\documentclass[12pt]{article}
\usepackage{amssymb,amsmath}
\usepackage[noblocks]{authblk}
\usepackage[top=0.75in, bottom=0.75in, left=0.75in, right=0.75in, dvips]{geometry}
\usepackage{caption}
\usepackage{graphicx}
\usepackage{epstopdf}
\date{}
\begin{document}
	\textwidth 10.0in
	\textheight 9.0in
	\topmargin -0.60in
	\title{Renormalization Scheme Ambiguities in the Models with More than One Coupling}
	\author[1,2]{D.G.C. McKeon}
	\author[1]{Chenguang Zhao}
	\affil[1] {Department of Applied Mathematics, The
		University of Western Ontario, London, ON N6A 5B7, Canada}
	\affil[2] {Department of Mathematics and
		Computer Science, Algoma University,\newline Sault Ste. Marie, ON P6A
		2G4, Canada}
	
	\maketitle                                 
	
	\maketitle
	\noindent
	email: dgmckeo2@uwo.ca, czhao52@uwo.ca

\begin{abstract}
	The process of renormalization to eliminate divergences arising in quantum field theory is not uniquely defined; one can always perform a finite renormalization, rendering finite perturbative results ambiguous. The consequences of making such finite renormalizations have been examined in the case of there being one or two couplings. In this paper we consider how finite renormalizations can affect more general models in which there are more than two couplings. In particular we consider the massless limit of the Standard Model in which there are essentially five couplings. We show that in this model (when neglecting all mass parameters) if we use mass independent renormalization, then the renormalization group $\beta$-functions are not unique beyond one loop order, that it is not in general possible to eliminate all terms beyond certain order for all these $\beta$-functions, but that for a physical process all contributions beyond one loop order can be subsumed into the $\beta$-functions.
\end{abstract}

\section{Introduction}
One of the difficulties in perturbative quantum field theory is that the process of renormalization (made necessary by the occurence of explicit divergences) is not unique; the results of any calculation done to finite order can be altered by a finite renormalization of the quantities that characterize the theory. This was in fact initially recognized by Stueckelberg and Peterman [1,2,3,4,5,6]. The ambiguity that has received the most attention is associated with the non-physical mass scale $\mu$ that arises in massless quantum chromodynamics (QCD), a theory with a single coupling. Only upon summing perturbative contributions to all orders in perturbation theory can ambiguities in $\mu$ be overcome [7,8,9,10]. There are, however, additional ambiguities in perturbative quantum field theory that are a consequence of being able to make finite renormalizations. When there is a single coupling as in QCD, it has been shown that when using mass independent renormalization [4,5] these additional ambiguities can be characterized by a the coefficients of a perturbative expansion of the function $\beta$ which controls how the coupling varies with $\mu$ [11]. However, an examination of the "renormalization scheme" (RS) ambiguities that occur when there are two couplings reveals quantitative differences with the RS ambiguities when there is but one coupling [10,12,13]. When there are two couplings, it is still possible to eliminate the dependence on $\mu$ by summing to all orders of perturbation theory, but now ambiguities associated with performing a finite renormalization cannot be characterized by the expansion coefficients of the $\beta$-function. It is possible though to choose a RS so that all radiative effects beyond one loop order do not contribute to the $\beta$ function of at least one of the couplings, or do not contribute to higher order radiative effects associated with the physical process being considered [10]. We now will examine the consequences of these renormalzation scheme ambiguities when there are more than just two couplings, such as in the Standard Model.

In its simplest form, the conformal limit of the Standard Model of particle physics involves five coupling constants $g_a$: the SU(3), SU(2) and U(1) gauge couplings as well as the quartic SU(2) scalar self coupling and the Yukawa coupling of the top quark. As with any renormalizable theory, renormalization introduces a mass scale $\mu$ and these couplings all vary as $\mu$ varies in a way dictated by the renormalizaion group (RG) $\beta$-functions:

\begin{align}\tag{1}
\mu \frac{dg_a}{d\mu}=\beta_a (g_b)=
\sum_{k=2}^\infty \sum_{i_1=0}^k \sum_{i_2=0}^k \sum_{i_3=0}^k \sum_{i_4=0}^k \sum_{i_5=0}^k c_{k;i_1 i_2 i_3 i_4 i_5}^a({g}_1)^{i_1} ({g}_2)^{i_2} ({g}_3)^{i_3} ({g}_4)^{i_4} ({g}_5)^{i_5} \delta_{k-(i_1+i_2+...+i_5)} .
\end{align}

where the expansion coefficients ${c}_{k;i_1 i_2 i_3 i_4 i_5}^a$ are computed explictly by evaluating the appropriate $k-1$ loop Feynman diagrams.

However, the value of any physical quantity $R$ when computed to finite order in perturbative theory has explicit dependence on $\mu$. This explicit dependence must be compensated for by the implicit dependence through $g_a(\mu)$; this leads to the RG equation [1-6]

\begin{align}\tag{2}
\mu \frac{d}{d\mu} R = (\mu \frac{\partial}{\partial\mu} + \beta_a (g_b) \frac{\partial}{\partial g_a}) R = 0.
\end{align}

In addition to the ambiguity in the perturbative value of $R$ resulting from the necessity of introducing the renormalization mass scale $\mu$, it is possible to make finite renormalizations of the couplings $g_a$, even when using a mass-dependent renormalization scheme (RS) [4,5], so that $g_a$ is replaced by $\overline{g}_a$ where

\begin{equation}\tag{3}
\overline{g}_a = g_a + 
\sum_{k=2}^\infty \sum_{i_1=0}^k \sum_{i_2=0}^k \sum_{i_3=0}^k \sum_{i_4=0}^k \sum_{i_5=0}^k x_{k;i_1 i_2 i_3 i_4 i_5}^a(g_1)^{i_1} (g_2)^{i_2} (g_3)^{i_3} (g_4)^{i_4} (g_5)^{i_5} \delta_{k-(i_1+i_2+...+i_5)}.
\end{equation}

with the expansion coefficients $x_{k;i_1 i_2 i_3 i_4 i_5}^a$ characterizing the change in RS.

There is an extensive literature dealing with the RS ambiguities when there is one coupling (for example ref [11]). These considerations have been extended to the case of two couplings [10]. There are qualitative differences between the RS ambiguities occurring when there are one and two couplings. When there is one coupling a, the RS ambiguities can be characterized by the coefficients of the $\beta$-function $\beta (a)$ [11] and a RS can be chosen so that $\beta (a)$ receives no contribution beyond two loop order [6,14]. Furthermore, it is possible to have a RS so that $R(a)$ vanishes beyond one-loop order and all higher loop effects serve only to affect the $\beta$-function [7,8]. This can be implemented after the RG equation (2) is used to sum all logarithmic contribution to $R$ which results in a cancellation between the implicit and explicit dependence on $\mu$ [7-10].

In ref [10], the consequences of there being RS ambiguities when there are two couplings are considered; in this paper we look at RS ambiguities in the conformal(massless) limit of the Standard Model where there are five couplings that are of importance. It turns out that many of the novel features associated with there being two couplings also occur when there are five couplings. In particular, the coefficients of the $\beta$-function can no longer be used to characterize the RS being used (as they are when there is one coupling [11]). We also find that just as in the two coupling case, when there are five couplings all radiative effects can be absorbed into the $\beta$-function coefficients beyond one-loop order so that the perturbative expansion for any physical quantity terminates at lowest order. A third feature of RS ambiguities when there are five couplings is that there is the possibility of working in a RS in which all contributions to at least one of the $\beta$-functions terminate at one-loop order, much as in the 't Hooft RS [14] for QCD. However, no matter how many couplings there are, if it is possible to sum all dependence on $ln\mu$ to all orders in perturbation theory, the explicit and implicit dependence on $\mu$ cancels so that this ambiguity in a perturbative calculation of a physical processes no longer presents a problem [7-10].

We will examine the effects of RS ambiguities on the couplings in the Standard Model when we restrict our attention to the three gauge couplings, a single four point Higgs coupling and the Yukawa coupling of the top quark. We note that when using modified minimal substraction $\overline{MS}$ as a RS, then all $\beta_a(g_b)$ have been computed to two loop order [15] while the  $\beta$-function for the gauge couplings are known to three loop order [16].

\section{The Standard Model}

We begin by noting that if a $\beta$-function $\overline{\beta}_a(\overline{g}_a)$ dictates how $\overline{g}_a$ evolves under change of $\mu$ and $\overline{\beta}_a(\overline{g}_a)$ has the same form as eq. (1) with $\overline{c}_{k;i_1 i_2 i_3 i_4 i_5}^a$ replacing $c_{k;i_1 i_2 i_3 i_4 i_5}^a$,  then since both

\begin{align}\tag{4}
\mu \frac{d\overline{g}_a}{d\mu} = \overline{\beta}_a (\overline{g}_b(g_c))
\end{align}

and 

\begin{align}\tag{5}
\mu \frac{d\overline{g}_a}{d\mu} = \sum_{c=1}^{5} \frac{\partial \overline{g}_a (g_c)}{\partial g_c} \beta_c (g_b)
\end{align}

where $\overline{g}_b(g_c)$ is given by eq. (3), we find from eqs. (4,5) that upon looking at terms quadratic and cubic in the couplings

\begin{align}\tag{6a}
\overline{c}_{2;i_1 i_2 i_3 i_4 i_5}^a
= {c}_{2;i_1 i_2 i_3 i_4 i_5}^a
\end{align}

\begin{align}\tag{6b}
\overline{c}_{3;30000}^1
& = c_{3;30000}^1 + c_{2;20000}^5 x_{2;10001}^1 + c_{2;20000}^4 x_{2;10010}^1 +c_{2;20000}^3 x_{2;10100}^1 + c_{2;20000}^2 x_{2;11000}^1 \\
& - c_{2;11000}^1 x_{2;20000}^2 - c_{2;10100}^1 x_{2;20000}^3 - c_{2;10010}^1 x_{2;20000}^4 - c_{2;10001}^1 x_{2;20000}^5 \nonumber
\end{align}

and

\begin{align}\tag{6c}
\overline{c}_{3;21000}^1
& = c_{3;21000}^1  + c_{2;20000}^5 x_{2;01001}^1 + c_{2;20000}^4 x_{2;01010}^1 + c_{2;20000}^3 x_{2;01100}^1 + 2 c_{2;20000}^2 x_{2;02000}^1 \\ \nonumber
& + c_{2;11000}^5 x_{2;10001}^1 + c_{2;11000}^4 x_{2;10010}^1 + c_{2;11000}^3 x_{2;10100}^1 + c_{2;11000}^2 x_{2;11000}^1 + c_{2;11000}^1 x_{2;20000}^1 \\ \nonumber
& - 2 c_{2;02000}^1 x_{2;20000}^2 - c_{2;01100}^1 x_{2;20000}^3 - c_{2;01010}^1 x_{2;20000}^4 - c_{2;01001}^1 x_{2;20000}^5 \\ \nonumber
& - c_{2;20000}^1 x_{2;11000}^1 - c_{2;11000}^1 x_{2;11000}^2 - c_{2;10100}^1 x_{2;11000}^3 - c_{2;10010}^1 x_{2;11000}^4 - c_{2;10001}^1 x_{2;11000}^5. \nonumber
\end{align}

etc.

If there were but one coupling, eq. (6) shows that $c_2$ and $c_3$ are unaltered by a change of RS of the form of eq. (3) [6]; $c_n (n>3)$  which arise from an (n-1) loop calculation can all be altered. In fact, $x_n (n>2)$  can be chosen so that $c_n (n>3)$ vanishes [14]. A RS can be characterized either by $c_n (n>3)$ [11] with $\mu$ being identified with $x_2$ [8], or by the parameters $x_n (n\geq 2)$ themselves [8].

It is possible to see that with five coupling constants, as with two coupling constants [10], there simply are not enough constants appearing in the expansion of $\overline{g}_a$ given in eq.(3) to independently vary the constants in the expansion of $\beta_a(g_b)$ in eq. (1). (In particular, at N-loop order, there are more constants $c_{N+1;i_1 i_2 i_3 i_4 i_5}^a$ than constants $x_{N;i_1 i_2 i_3 i_4 i_5}^a$.) Thus, unlike what happens when there is one coupling, the coefficients of the expansion of $\beta_a(g_b)$ are not suitable for characterizing a RS and as in the case of two couplings, we will employ directly the coefficients $x_{N;i_1 i_2 i_3 i_4 i_5}^a$ of eq. (3) to relate the parameters that occur when using a particular RS to that of a "base scheme", such as minimal subtraction (MS) [5].

In particular, since

\begin{equation}\tag{7}
\frac{\partial\overline{g}_a}{\partial x_{k;i_1 i_2 i_3 i_4 i_5}^b} = \overline{B}_{b;k;i_1 i_2 i_3 i_4 i_5}^a (\overline{g}_c) = \delta_b^a \delta_{k-(i_1+i_2+...+i_5)} g_1^{i_1}g_2^{i_2}g_3^{i_3}g_4^{i_4}g_5^{i_5}.
\end{equation}

and as eq. (3) can be inverted to give

\begin{align}\tag{8}
g_a & = \overline{g}_a + 
\sum_{k=2}^\infty \sum_{i_1=0}^k \sum_{i_2=0}^k \sum_{i_3=0}^k \sum_{i_4=0}^k \sum_{i_5=0}^k y_{k;i_1 i_2 i_3 i_4 i_5}^a(\overline{g}_1)^{i_1} (\overline{g}_2)^{i_2} (\overline{g}_3)^{i_3} (\overline{g}_4)^{i_4} (\overline{g}_5)^{i_5} \delta_{k-(i_1+i_2+...+i_5)} 
\end{align}

where

\begin{equation}\tag{9}
y_{2;i_1 i_2 i_3 i_4 i_5}^a + x_{2;i_1 i_2 i_3 i_4 i_5}^a = 0 \:(a=1,2,...5;i_1+i_2+i_3+i_4+i_5=2)
\end{equation}

\begin{subequations}
	\begin{align}\tag{10a}
	y_{3;30000}^1 &= -x_{3;30000}^1+2(x_{2;20000}^1)^2 + x_{2;11000}^1 x_{2;20000}^2  \nonumber \\
	& + x_{2;10100}^1 x_{2;20000}^3 + x_{2;10010}^1 x_{2;20000}^4 + x_{2;10001}^1 x_{2;20000}^5 \nonumber \\ \tag{10b}
	y_{3;21000}^1 &= -x_{3;21000}^1+3(x_{2;20000}^1)^2 + x_{2;11000}^1 x_{2;20000}^2  \nonumber \\
	& + x_{2;10100}^1 x_{2;20000}^3 + x_{2;10010}^1 x_{2;20000}^4 + x_{2;10001}^1 x_{2;20000}^5 \nonumber \\ \tag{10c}
	y_{3;11100}^1 &= -x_{3;11100}^1 + 2x_{2;10100}^1 x_{2;11000}^1 + 2x_{2;01100}^1 x_{2;20000}^1  \nonumber \\
	& + x_{2;11000}^1 x_{2;01100}^2 + 2x_{2;02000}^1 x_{2;10100}^2 + x_{2;01100}^1 x_{2;11000}^2  \nonumber \\
	& + x_{2;10100}^1 x_{2;01100}^3 + x_{2;01100}^1 x_{2;10100}^3 + 2x_{2;00200}^1 x_{2;11000}^3  \nonumber \\
	& + x_{2;10010}^1 x_{2;01100}^4 + x_{2;01010}^1 x_{2;10100}^4 + x_{2;00110}^1 x_{2;11000}^4  \nonumber \\
	& + x_{2;10001}^1 x_{2;01100}^5 + x_{2;01001}^1 x_{2;10100}^5 + x_{2;00101}^1 x_{2;11000}^5 \nonumber
	\end{align}
\end{subequations}

etc.

we find that eq. (7) leads to, for example

\begin{align}\tag{11}
\frac{d\overline{g}_1}{dx_{2;02000}^1} = \overline{B}_{1;2;02000}^1 (\overline{g}_c) = g_2^2 = \overline{g}_2^2 &- 
x_{2;20000}^2 \overline{g}_1^2 \overline{g}_2 - x_{2;02000}^2 \overline{g}_2^3 - 
x_{2;00200}^2 \overline{g}_3^2 \overline{g}_2 - x_{2;00020}^2 \overline{g}_4^2 \overline{g}_2 \\ 
& - x_{2;00002}^2 \overline{g}_5^2 \overline{g}_2 - x_{2;11000}^2 \overline{g}_1 \overline{g}_2^2 - x_{2;01100}^2 \overline{g}_2 \overline{g}_3^2 - x_{2;00110}^2 \overline{g}_2 \overline{g}_3 \overline{g}_4 \ldots\nonumber
\end{align}

As noted above, in ref [9] it is shown that if there is one coupling, there exists a RS in which $c_n=0$ beyond two loop order. In contrast, by eq. (6) we cannot find a scheme when there are five couplings such that $c_{k;i_1 i_2 i_3 i_4 i_5}^a$ all vanish beyond a certain order in the loop expansion. However, it is possible to find a RS in which at least one of the couplings has a $\beta$-function that receives no contribution beyond one loop order. For example, if $x_{k;i_1 i_2 i_3 i_4 i_5}^a=0 (a\neq 1)$ then eqs. (6a-c) simplify and we obtain the relations between $\overline{c}_{3;i_1 i_2 i_3 i_4 i_5}^a$ and $c_{3;i_1 i_2 i_3 i_4 i_5}^a$

\begin{align}\tag{12a}
\overline{c}_{3;30000}^1
& = c_{3;30000}^1 + c_{2;20000}^5 x_{2;10001}^1 + c_{2;20000}^4 x_{2;10010}^1 +c_{2;20000}^3 x_{2;10100}^1 + c_{2;20000}^2 x_{2;11000}^1 
\end{align}

\begin{align}\tag{12b}
\overline{c}_{3;21000}^1
& = c_{3;21000}^1  + c_{2;20000}^5 x_{2;01001}^1 + c_{2;20000}^4 x_{2;01010}^1 + c_{2;20000}^3 x_{2;01100}^1 + 2 c_{2;20000}^2 x_{2;02000}^1 \\ \nonumber
& + c_{2;11000}^5 x_{2;10001}^1 + c_{2;11000}^4 x_{2;10010}^1 + c_{2;11000}^3 x_{2;10100}^1 + c_{2;11000}^2 x_{2;11000}^1  \\ \nonumber
& + c_{2;11000}^1 x_{2;20000}^1 - c_{2;20000}^1 x_{2;11000}^1 \nonumber
\end{align}

etc. and

\begin{align}\tag{13a}
\overline{c}_{3;30000}^2
& = c_{3;30000}^2 - 2 c_{2;20000}^2 x_{2;20000}^1
\end{align}

\begin{align}\tag{13b}
\overline{c}_{3;21000}^2
& = c_{3;21000}^2  -2 c_{2;20000}^2 x_{2;11000}^1 - c_{2;11000}^2 x_{2;20000}^1 
\end{align}

\begin{align}\tag{13c}
\overline{c}_{3;11100}^2
& = c_{3;21000}^2  -2 c_{2;20000}^2 x_{2;01100}^1 - c_{2;11000}^2 x_{2;10100}^1 - c_{2;10100}^2 x_{2;11000}^1 .
\end{align}

etc.

with all other $\overline{c}_{k;i_1 i_2 i_3 i_4 i_5}^a$ similarly computed. We see that it is possible to choose $x_{k;i_1 i_2 i_3 i_4 i_5}^1$ so that $c_{k;i_1 i_2 i_3 i_4 i_5}^1=0$ for all $k>2$. We could, for example, identify $g_1$ with the strong SU(3) coupling $16\pi^2 a$ in which case $\overline{a}$ would by ref. [12] satisfy simply

\begin{align}\tag{14}
\mu \frac{d\overline{a}}{d\mu} = -14 \overline{a}^2
\end{align}

with no higher loop corrections. Of course, in this scheme, $\overline{g}_2 ... \overline{g}_5$ would all satisfy eq. (1) with coefficients $\overline{c}_{k;i_1 i_2 i_3 i_4 i_5}^a (a=2,3,4,5)$ that depend on the values of $x_{k;i_1 i_2 i_3 i_4 i_5}^a$ chosen to give rise to eq. (14). Limiting the number of terms that contribute to a $\beta$-function, as in eq. (14), is the multi-coupling analogue of using the 't Hooft RS [14] when there is but one coupling.

We now will consider RS dependence for a physical quantity $R$ expanded as 

\begin{equation}\tag{15}
R = \sum_{k=0}^\infty A_k(a) L^k
\end{equation}

where $L = \ln (\frac{\mu}{Q})$ and
\begin{equation}\tag{16}
A_k(a) = \sum_{m=0}^\infty \sum_{i_1=0}^\infty \sum_{i_2=0}^\infty ... \sum_{i_5=0}^\infty T_{m;i_1 i_2 ... i_5;k} \delta_{m+k+1-i_1-i_2...-i_5} (g_1)^{g_1} (g_2)^{g_2} ... (g_5)^{g_5}.
\end{equation}

With $g_2$ satisfying eq. (1), substitution of eq. (15) into eq.(2) leads to

\begin{equation}\tag{17}
A_{k+1}(g_a(\ln (\frac{\mu}{\Lambda}))) = \frac{-1}{k+1} \frac{d}{d(\ln \frac{\mu}{\Lambda})} A_{k}(g_a(\ln \frac{\mu}{\Lambda}))
\end{equation}

where $\Lambda$ is a mass scale associated with the boundary conditions on eq. (1). As a result [7]

\begin{equation}\tag{18}
R = \sum_{k=0}^\infty \frac{(-L)^k}{k!} (\frac{d}{d(\ln \frac{\mu}{\Lambda})})^k A_{0}(g_a(\ln \frac{\mu}{\Lambda})) = A_{0}(g_a(\ln \frac{Q}{\Lambda})).
\end{equation}

All explicit dependence of $R$ on $\mu$ through L in eq. (15) has been canceled with the implicit dependence on $\mu$ through $g_a(\ln (\frac{\mu}{\Lambda}))$ upon summing the logarithmic terms in eq. (15), which is possible on account of the RG equation (2). The apparent ambiguity in the perturbative expansion for $R$ due to $\mu$ has disappeared.

Together eqs. (16) and (18) lead to

\begin{equation}\tag{19}
R = \sum_{m=0}^\infty \sum_{i_1=0}^\infty \sum_{i_2=0}^\infty ... \sum_{i_5=0}^\infty T_{m;i_1 i_2 ... i_5} \delta_{m+1-i_1-i_2...-i_5} (g_1)^{i_1} (g_2)^{i_2} ... (g_5)^{i_5}
\end{equation}

where 

\begin{equation}\tag{20}
T_{m;i_1 i_2 ... i_5} = T_{m;i_1 i_2 ... i_5;0}.
\end{equation}

Under the change in RS in eq. (3), we have $T$ and $g_a$ in eq. (18) replaced by $\overline{T}$ and $\overline{g}_a$. However, as $R$ is RS independent, we must have by eq. (7)

\begin{align}\tag{21}
\frac{dR}{dx_{k;i_1 i_2 i_3 i_4 i_5}^a} = 0 =& \left( \frac{\partial}{\partial x_{k;i_1 i_2 i_3 i_4 i_5}^a} + \overline{B}_{a;k;i_1 i_2 i_3 i_4 i_5}^b (\overline{g}_b)  \frac{\partial}{\partial \overline{g}^b}\right) \nonumber \\
&\sum_{h=1}^\infty \sum_{j_1=0}^h \sum_{j_2=0}^h \sum_{j_3=0}^h \sum_{j_4=0}^h \sum_{j_5=0}^h \delta_{h-(j_1+j_2+...+j_5)}
\overline{T}_{h;j_1 j_2 j_3 j_4 j_5} (\overline{g}_1)^{j_1} (\overline{g}_2)^{j_2}(\overline{g}_3)^{j_3}(\overline{g}_4)^{j_4}(\overline{g}_5)^{j_5} . \nonumber
\end{align}

Upon using eq. (7) for $\overline{B}_{a;k;i_1 i_2 i_3 i_4 i_5}^b$, eq. (21) becomes

\begin{align}\tag{22}
=\sum_{h=1}^\infty \sum_{j_1=0}^h \sum_{j_2=0}^h \sum_{j_3=0}^h \sum_{j_4=0}^h \sum_{j_5=0}^h & \delta_{h-(j_1+j_2+...+j_5)} \Bigg\{ \frac{\partial \overline{T}_{h;j_1 j_2 j_3 j_4 j_5}}{\partial x_{k;i_1 i_2 i_3 i_4 i_5}^a} (\overline{g}_1)^{j_1} (\overline{g}_2)^{j_2}(\overline{g}_3)^{j_3}(\overline{g}_4)^{j_4}(\overline{g}_5)^{j_5} \nonumber \\
& + \overline{T}_{h;j_1 j_2 j_3 j_4 j_5} \Big[ \overline{B}_{a;k;i_1 i_2 i_3 i_4 i_5}^1 j_1
(\overline{g}_1)^{j_1-1} (\overline{g}_2)^{j_2}(\overline{g}_3)^{j_3}(\overline{g}_4)^{j_4}(\overline{g}_5)^{j_5} \nonumber \\
&+ \overline{B}_{a;k;i_1 i_2 i_3 i_4 i_5}^2 j_2
(\overline{g}_1)^{j_1} (\overline{g}_2)^{j_2-1}(\overline{g}_3)^{j_3}(\overline{g}_4)^{j_4}(\overline{g}_5)^{j_5} +... \Big]\Bigg\} = 0.\nonumber
\end{align}

Terms of a given order in $\overline{g}_a$ lead to, for example

\begin{equation}\tag{23}
\frac{\partial \overline{T}_{1;j_1 j_2 j_3 j_4 j_5}}{\partial x_{k;i_1 i_2 i_3 i_4 i_5}^a} = 0 
\end{equation}
\begin{equation}\tag{24a}
\frac{\partial \overline{T}_{2;20000}}{\partial x_{k;i_1 i_2 i_3 i_4 i_5}^a} + \left( \overline{T}_{1;10000} \delta_1^a + \overline{T}_{1;01000} \delta_2^a + \overline{T}_{1;00100} \delta_3^a + \overline{T}_{1;00010} \delta_4^a +  \overline{T}_{1;00001} \delta_5^a \right) \delta_{j_1 2} \delta_{j_2 0} \delta_{j_3 0} \delta_{j_4 0} \delta_{j_5 0} = 0
\end{equation}
\begin{equation}\tag{24b}
\frac{\partial \overline{T}_{2;11000}}{\partial x_{k;i_1 i_2 i_3 i_4 i_5}^a} + \left( \overline{T}_{1;10000} \delta_1^a + \overline{T}_{1;01000} \delta_2^a + \overline{T}_{1;00100} \delta_3^a + \overline{T}_{1;00010} \delta_4^a +  \overline{T}_{1;00001} \delta_5^a \right) \delta_{j_1 1} \delta_{j_2 1} \delta_{j_3 0} \delta_{j_4 0} \delta_{j_5 0} = 0
\end{equation}
\begin{equation}\tag{24c}
\frac{\partial \overline{T}_{2;02000}}{\partial x_{k;i_1 i_2 i_3 i_4 i_5}^a} + \left( \overline{T}_{1;10000} \delta_1^a + \overline{T}_{1;01000} \delta_2^a + \overline{T}_{1;00100} \delta_3^a + \overline{T}_{1;00010} \delta_4^a +  \overline{T}_{1;00001} \delta_5^a \right) \delta_{j_1 0} \delta_{j_2 2} \delta_{j_3 0} \delta_{j_4 0} \delta_{j_5 0} = 0
\end{equation}
etc. \\

These equations have the boundary conditions that $\overline{T}=T$ when $x_{k;i_1 i_2 i_3 i_4 i_5}^a=0$ and so we have the solutions

\begin{equation}\tag{25}
\overline{T}_{1;j_1 j_2 j_3 j_4 j_5} = T_{1;j_1 j_2 j_3 j_4 j_5}
\end{equation}
\begin{equation}\tag{26a}
\overline{T}_{2;20000} = T_{2;20000} - x_{2;20000}^1 T_{1;10000} - x_{2;20000}^2 T_{1;01000} - x_{2;20000}^3 T_{1;00100} - x_{2;20000}^4 T_{1;00010} - x_{2;20000}^5 T_{1;00001}
\end{equation}
\begin{equation}\tag{26b}
\overline{T}_{2;11000} = T_{2;11000} - x_{2;11000}^1 T_{1;10000} - x_{2;11000}^2 T_{1;01000} - x_{2;11000}^3 T_{1;00100} - x_{2;11000}^4 T_{1;00010} - x_{2;11000}^5 T_{1;00001}
\end{equation}
\begin{equation}\tag{26c}
\overline{T}_{2;02000} = T_{2;02000} - x_{2;02000}^1 T_{1;10000} - x_{2;02000}^2 T_{1;01000} - x_{2;02000}^3 T_{1;00100} - x_{2;02000}^4 T_{1;00010} - x_{2;02000}^5 T_{1;00001}
\end{equation}
etc.\\

It is evident that these equations can be used to find values of $x_{k;i_1 i_2 i_3 i_4 i_5}^a$ that lead to $\overline{T}_{m;j_1 j_2 j_3 j_4 j_5}=0$ with $m\geq 2$. In this case we have 

\begin{equation}\tag{27}
R = T_{1;10000} \overline{g}_1\left(\ln \frac{Q}{\Lambda}\right) + T_{1;01000} 
\overline{g}_2\left(\ln \frac{Q}{\Lambda}\right) + T_{1;00100} \overline{g}_3\left(\ln \frac{Q}{\Lambda}\right) + T_{1;00010} \overline{g}_4\left(\ln \frac{Q}{\Lambda}\right) + T_{1;00001} \overline{g}_5\left(\ln \frac{Q}{\Lambda}\right)
\end{equation}

and no higher powers of $\overline{g}_a$ contribute to $R$. The values of $x_{k;i_1 i_2 i_3 i_4 i_5}^a$ that lead to eq. (27) are not unique. For example, if we choose to have $x_{h;i_1 i_2 i_3 i_4 i_5}^a=0$ for $a\neq 1$, then $\overline{T}_{m;j_1 j_2 j_3 j_4 j_5}=0 (m>1)$ results in

\begin{equation}\tag{28}
x_{2;i_1 i_2 i_3 i_4 i_5}^1 = \frac{T_{2;i_1 i_2 i_3 i_4 i_5}}{T_{1;10000}}
\end{equation}
\begin{equation}\tag{29a}
x_{3;30000}^1 = \frac{2(T_{2;20000})^2}{(T_{1;10000})^2} + \frac{T_{3;30000}}{T_{1;10000}}
\end{equation}
\begin{equation}\tag{29b}
x_{3;12000}^1 = \frac{(T_{2;11000})^2+T_{2;02000}T_{2;20000}}{(T_{1;10000})^2} + \frac{T_{3;12000}}{T_{1;10000}}
\end{equation}
\begin{equation}\tag{29c}
x_{3;11100}^1 = \frac{T_{2;10100}T_{2;11000}+T_{2;01100}T_{2;20000}}{(T_{1;10000})^2} + \frac{T_{3;11100}}{T_{1;10000}}
\end{equation}
etc.\\

The $\beta$-functions associated with $\overline{g}_a$ are now given by eqs. (12,13) with $x_{m;i_1 i_2 i_3 i_4 i_5}^a=0 (m=2,3)$ given by eq. (28,29).

These general equations can be applied to the analysis of the specific physical processes, such as the total hadronic cross section in electron-positron annihilation or the decay of the Higgs Boson.

\section{Discussion}

We have demonstrated that the possibility of making a finite renormalization of the five couplings provides a great deal of flexibility in the way perturbative results can be presented. It is possible to reduce the $\beta$-function for one of the couplings to the one loop result. It is also possible to sum all logarithmic contributions to a physical quantity $R$, thereby eliminating dependence on the renormalization mass scale $\mu$ and to make it possible to eliminate all higher order contributions to $R$. Any finite renormalization serves to affect the contributions to the $\beta$-functions beyond one loop order. 

We plan to examine the RS ambiguities in non-conformal models in which there are both several masses and several couplings. Already these ambiguities have been considered when computing the semi-leptonic decay of the b quark [9]. In this process, there is but one mass (that of the b quark) and one coupling (the strong coupling).  

\section*{Acknowledgements}
\noindent
Roger Macleod had a useful suggestion.

\end{document}